\documentclass[twocolumn,amsmath,amssymb,nofootinbib,eqsecnum,tighten]{revtex4}

\usepackage{fancybox}
\usepackage{graphicx}
\usepackage{dcolumn} 
\usepackage{bm}      
\usepackage[dvips]{color} 
\usepackage{curves}

\begin{document}

\title{
Entanglement entropy
and the Berry phase in
solid states
}
\author{S.\ Ryu}
\affiliation{Kavli Institute for Theoretical Physics,
	     University of California, 
	     Santa Barbara, 
	     CA 93106, 
	     USA}
\author{Y.\ Hatsugai}
\affiliation{Department of Applied Physics,
             University of Tokyo,
             Hongo Bunkyo-ku,
             Tokyo 113-8656,
             Japan}
\date{\today}

\begin{abstract}
The entanglement entropy (von Neumann entropy)
has been used to characterize the complexity
of many-body ground states in strongly correlated systems.
In this paper, we try to establish a connection between
the lower bound of the von Neumann entropy and the
Berry phase defined for quantum ground states.
As an example, 
a family of translational invariant lattice free fermion
systems
with two bands separated by a finite gap
is investigated.
We argue that, for one dimensional (1D) cases, 
 when the Berry phase (Zak's phase) of 
the occupied band is 
equal to $\pi \times (\mbox{odd integer})$ 
and when the ground state respects a discrete unitary
particle-hole symmetry (chiral symmetry), 
the entanglement entropy in the thermodynamic limit 
is at least larger than $\ln 2$ (per boundary),
i.e., the entanglement entropy that corresponds 
to a maximally entangled pair of two qubits.
We also discuss this lower bound 
is related to vanishing of the expectation
value of a certain non-local operator which creates a kink
in 1D systems. 
\end{abstract}

\maketitle


\section{Introduction}
\label{sec: introduction}

One of the most distinctive features of
quantum phases of matter is that they
are not completely
characterized by their pattern of symmetry breaking
(order parameters of some kind),
which is in a sharp contrast to classical statistical systems.
Instead, quantum ground states
should be described by
their pattern of entanglement such as
topological or quantum order.
\cite{Wen89}
However, beyond some simple textbook examples, e.g., 
a system of two coupled $S=1/2$ spins (qubits),
we do not have many intuitions about quantum entanglement
hidden in many-body wave functions.
In a recent couple of years, 
the entropy of entanglement (von Neumann entropy) 
\cite{footnote1}
\begin{eqnarray}
S_A = - \mathrm{tr}_{A}\, \rho_{A} \ln \rho_{A},\quad
\rho_{A}=\mathrm{tr}_{B}\, |\Psi\rangle \langle \Psi|,
\label{eq: def entanglement entropy}
\end{eqnarray}
has been used to measure
how closely entangled (or how ``quantum'')
a given ground state wave function $|\Psi\rangle$ is.
Here,
the total system is divided
into two subsystems $A$ and $B$ and 
$\rho_{A}$ is the reduced density matrix 
for the subsystem $A$
obtained by taking a partial trace over the subsystem $B$
of the total density matrix $\rho=|\Psi\rangle \langle \Psi|$.
This quantity 
is zero for classical product states
whereas 
it takes a non-trivial value
for valence-bond solid states (VBS),
or resonating valence bond states (RVB)
of quantum spin systems, say.
Recently,
the entanglement entropy
at and close to quantum phase transitions
in low-dimensional strongly correlated
systems has been used as a new tool
to investigate the nature of quantum criticality.
\cite{
Holzhey94,
Osterloh02,
Osborne02,
Vidal03,
Fan04,
Calabrese04,
Furukawa05}
Even though one can tell different quantum phases from
the scaling of the entanglement entropy,
it is still not completely understood
what kind of information
we can distill from the von Neumann entropy,
other than that contained in 
conventional correlation functions.

On the other hand, a phase degree of freedom is also
a specific feature of quantum mechanics.
Indeed, Berry phases \cite{Berry84}
associated with
(many-body) wave functions in solid states
are related to several interesting quantum phenomena
which have no classical analogue.
Probably, it is best epitomized by 
the Thouless-Kohmoto-Nightingale-Nijs (TKNN)
formula in the integer quantum Hall effect (IQHE)
\cite{Thouless82,kohmoto},
in which gapped quantum phases are distinguished 
by an integral topological invariant originating from 
winding of the phase of wave functions.
In addition to the IQHE,
the Berry phase also appears 
in the King-Smith-Vanderbilt (KSV) formula 
\cite{Kingsmith93, Resta94}
of the theory of macroscopic polarization,
and its incarnation in quantum spin chains
\cite{denNijs89,Nakamura02},
and so on.
An observable consequence of the non-trivial Berry phase
is the existence of localized states
at the boundaries 
when we terminate a system with boundaries.
\cite{Hatsugai93,Ryu02,Kitaev00}


It is then tempting to ask if there is, 
if any, a connection between these two
paradigms in quantum physics, namely,
entanglement and the Berry phase.
In this paper, we discuss this issue by taking a family of 
translational invariant lattice 
free fermion systems in $d$ dimensions
as an example. We bipartition the system into
two subsystems $A$ and $B$ by introducing 
$(d-1)$-dimensional flat interfaces. 
Within this setup, 
we can reduce the calculation of the entropy
to that in a one-dimensional system by 
the $(d-1)$-dimensional Fourier transform
along the interface.
We assume the existence of a finite energy gap $m$
above ground states
which is inversely proportional to the correlation length, 
$m \sim \xi_{corr}^{-1}$
(when measured in the unit of the band width).
Furthermore,
for simplicity, we consider the case in which
there are only two bands that are separated by a gap.

In this paper, we consider 
the Berry phase associated with 
a response of a quantum ground state to a continuous twist of the boundary condition.
For the case of free lattice fermion systems, 
for which a ground state is given by a filled Fermi-Dirac sea,
this Berry phase is a phase acquired by an adiabatic transport 
of the Bloch wave functions in the momentum space
and also called Zak's phase.\cite{Zak89}
Physically,
it is related to macroscopic polarization
of the Fermi-Dirac sea.
\cite{Kingsmith93}
A beauty of the simple two-band example that we discuss
is that the Berry phase for the quantum ground state
can be easily computed and visualized,
following the pioneering work by Berry
\cite{Berry84}
(See Sec.\ \ref{sec: 1D two-band systems} and Fig.\ \ref{fig:
bloch_sphere} below.)


With these setups, we will demonstrate that
taking the partial trace over a subsystem corresponds to 
creating boundaries in a system.
Two contributions to the entanglement entropy will be then identified.
The first one is of type already discussed in
a flurry of recent works focusing on detection of quantum critical
points.
This contribution to the entanglement entropy is largely controlled by
the correlation length $\xi_{corr}$.
For example,
in one-dimensional (1D) many-body systems close to criticality
the entanglement entropy 
obeys a logarithmic law 
$S_A \sim \mathcal{A}(c/6)\ln \xi_{corr}/a$
where 
$c$ is the central charge of the conformal field theory
that governs the criticality, 
$a$ the lattice constant,
and $\mathcal{A}$ is the the number of
boundary points of $A$.
\cite{Vidal03,Calabrese04}

On the other hand,
the second contribution to the entropy 
comes from 
the localized boundary states of the correlation matrix that 
exist when the Berry phase of the ground state wave function
is non-vanishing.
Especially, when the Berry phase is equal to $\pi \times \mbox{(odd integer)}$ 
and when the ground state respects 
discrete symmetries of some sort,
the localized boundary states are 
topologically protected 
as discussed in Refs.\ \cite{Hatsugai93,Ryu02}.
For this case, we will show that the contribution from the 
boundary states to the von Neumann entropy is $\ln 2$ per boundary,
i.e., the same amount of entropy carried by maximally entangled 
pair of two qubits.
We will also illustrate, by taking a specific limit,
that when $\gamma\neq 0$,
the von Neumann entropy from the boundary states
is that of partially entangled qubits.

We also discuss that the $\ln 2$ contribution to
the von Neumann entropy
is related to vanishing of the expectation
value of a certain non-local operator which creates a kink
in 1D systems.
This connection between the entanglement entropy and the kink operator is,
in flavor, similar to discussions in Refs.\ \cite{Calabrese04,Casini05} 
in which the entanglement entropy is expressed as the expectation values of 
twist operators
in conformal field theories.


The rest of the paper is organized as follows.
In Sec.\ \ref{sec: 1D two-band systems},
we start our discussions with 
1D translational invariant Hamiltonians
with two bands separated by a finite gap.
The Berry phase is introduced as
an expectation value of a specific non-local operator
that twists the phase of wavefunctions.
We then discuss its connection to
the entanglement entropy by making use of the 
correlation matrix.
The calculation of the entanglement entropy is,
in general, a rather difficult task at least analytically.
Furthermore, the Berry phase contribution
to the entropy might not be of perturbative nature. 
We thus consider two limiting situations.
In Subsec.\ \ref{subsec: limit 1},
we take the limit of the small correlation length $\xi_{corr} \ll 1$
and zero band width. In this specific limit, we can express
the entanglement entropy as a function of the Berry phase $\gamma$.
We next focus on cases with 
a discrete unitary particle-hole symmetry (chiral symmetry)
in Subsec.\ \ref{subsec: limit 2}.
Except for requiring the chiral symmetry,
any parameters of the Hamiltonian (the band structure) can be arbitrary.
Once we impose the chiral symmetry, 
the Berry phase $\gamma$ can take only discrete values,
integral multiple of $\pi$. 
We then show when $\gamma=\pi \times (\mbox{odd integer})$,
the entanglement entropy is bounded below as $S_A \ge 2 \ln 2$.
In Sec.\ \ref{sec: connection to  a kink operator},
we relate the lower bound of the entropy at $\gamma=\pi \times (\mbox{odd integer})$
to the vanishing of a non-local operator that creates a kink.
In Sec.\ \ref{sec: 2D systems with the non-vanishing Chern number},
these discussions are applied to a higher dimensional example, 
a 2D superconductor with non-zero TKNN integer.
We conclude in Sec.\ \ref{sec: conclusion}.

\section{1D two-band systems}
\label{sec: 1D two-band systems}


We start from the following 
1D translational invariant Hamiltonians
with two bands separated by a finite gap,
\begin{eqnarray}
\mathcal{H}
=
\sum_{x,x'}^{\mbox{\tiny PBC}}
\boldsymbol{c}_{x}^{\dag}
H_{x-x'}
\boldsymbol{c}_{x'}^{\ },
\quad
H_{x-x'}
=
\left(
\begin{array}{cc}
t_+ & \Delta \\
\Delta' & t_-
\end{array}
\right)_{x-x'}.
\label{eq: def 2-band 1D hamiltonian}
\end{eqnarray}
Here, a pair of fermion annihilation operators
$\boldsymbol{c}^{\mathrm{T}}_x=
(c_{+}^{\ }, c_{-}^{\ })_{x}$
is assigned for each cite,
$x,x'=1,\cdots, N$,
and 
the hermiticity of $\mathcal{H}$ implies
$t_{\iota,x-x'}=t_{\iota, x'-x}^*$
and
$\Delta_{x-x'}=(\Delta'_{x'-x})^*$
for $\iota=\pm$.
We impose the periodic boundary condition (PBC) on
the 1D lattice.
In spite of its simplicity,
this Hamiltonian (\ref{eq: def 2-band 1D hamiltonian})
has a wide range of applicability,
such as 
the Bogoliubov-de Gennes Hamiltonian in 
superconductivity, graphite systems \cite{Ryu02},
ferroelectricity of organic materials
and perovskite oxides \cite{Onoda04}, 
and the slave boson mean field theory for spin liquid states,
say.

By the Fourier transformation 
$
\boldsymbol{c}_x^{\ }
=
N^{-1/2}
\sum_{k \in \mathrm{Bz}}
e^{\mathrm{i}k x}
\boldsymbol{c}_{k}^{\ }
$
where the summation over $k$ extends over
the 1st Brillouin zone (Bz),
$k=2\pi n/N$ ($n=1,\ldots, N$),
the Hamiltonian in the momentum space is given by
$\mathcal{H}
=
\sum_{k \in \mathrm{Bz}}
\boldsymbol{c}^{\dag}_{k}
H(k)
\boldsymbol{c}^{\ }_{k}
$,
with
$
H(k)
:=
\sum_{x} e^{-\mathrm{i}k x}
H(x).
$
If we introduce an ``off-shell'' four-vector
$R^{\mu=0,1,2,3}(k)\in \mathbb{R}$ by
$
R^0(k)\mp R^3(k):= t_{\pm}(k)
$,
$
-R^1(k)+\mathrm{i}R^2(k) := \Delta (k)
$,
we can rewrite the Hamiltonian in the momentum space as
\begin{eqnarray}
\mathcal{H}
&=&
\sum_{k \in \mathrm{Bz}}
\boldsymbol{c}_{k}^{\dag}
R^{\mu}(k)\sigma_{\mu}
\boldsymbol{c}_{k}^{\ },
\label{eq: def 2-band 1D hamiltonian in k-space}
\end{eqnarray}
where 
$\boldsymbol{\sigma}_{\mu}=(\sigma_0,-\boldsymbol{\sigma})$
with $\sigma_0=\mathbb{I}_2$.

Observing that 
$R^{\mu}\sigma_{\mu}$ is diagonalized by
the same eigen vectors as 
those of $R_{i}\sigma_i=\boldsymbol{R}\cdot \boldsymbol{\sigma}$
(but with different eigen values),
normalized eigen states $\vec{v}_{\pm}$
for $R^{\mu}\sigma_{\mu}$
are given by, when $\boldsymbol{R}$ is not in the Dirac string,
$(R^1,R^2)\neq (0,0)$,
\cite{Berry84}
\begin{eqnarray}
\vec{v}_{\pm}
&=&
\frac{1}{\sqrt{2R(R\mp R^3)}}
\left(
\begin{array}{c}
R^1 -\mathrm{i} R^2 \\
\pm R -R^3
\end{array}
\right),
\label{eq : wave function for monopole}
\end{eqnarray}
where $R=|\boldsymbol{R}|$ (should not be confused with $R^0$),
and the eigen value for $\vec{v}_{\pm}$ is $E_{\pm}=R^0 \mp R$.
The Hamiltonian is then diagonalized as
$
\mathcal{H}
=
\sum_{k}
\boldsymbol{\alpha}_{k}^{\dag}
\mathrm{diag}(E_+,E_-)_{k}
\boldsymbol{\alpha}^{\ }_{k},
$
where
$
c_{\iota,k}
=
(\vec{v}_{\sigma})^{\iota}
\alpha_{\sigma,k}
$.
As we assume there is a finite gap 
for the entire Brillouin zone,
 $E_+ > E_-$, ${}^{\forall} k \in\mathrm{Bz}$.
The vacuum $|\Psi \rangle$ is the filled Fermi sea
$
|\Psi\rangle
=
\prod_{k \in \mathrm{Bz}}
\alpha_{-,k}^{\dag}
|0\rangle.
$

The Berry phase can be defined through
the expectation value of the twist operator:
\begin{eqnarray}
z &:=&
\exp\left[
\mathrm{i}\frac{2\pi}{N}\sum_{x}x n_{x}
\right],
\label{eq: expectation value of the twist operator}
\end{eqnarray}
where $n_x$ is the electron number operator at site $x$,
$n_{x}=\sum_{\iota}c^{\dag}_{x,\iota}c^{\ }_{x,\iota}$.
This operator twists the phase of wave functions along the $x$-direction
over the wide length scale, $N$.
If we use the $S_{z}$ component of spin operator, say, instead of
$n_x$, we can define
the twist operator in spin systems in a similar fashion.
The twist operator has been used to 
to characterize low-dimensional quantum systems
\cite{Nakamura02}
and to describe macroscopic polarization
of insulators
\cite{Kingsmith93}, say.

\begin{figure}
\begin{center}
\includegraphics[width=8cm,clip]{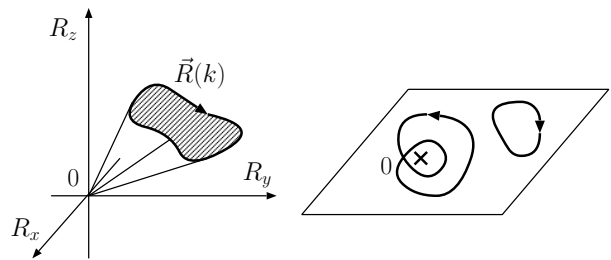}
\caption{
\label{fig: bloch_sphere}
(Left)
The loop defined by a three components vector $\boldsymbol{R}(k)$ 
associated with the Hamiltonian in momentum space
[Eq.\ (\ref{eq: def 2-band 1D hamiltonian})].
(Right)
Loops for chiral symmetric Hamiltonians.
}
\end{center}
\end{figure}

For the Fermi-sea $|\Psi\rangle$,
the expectation value of the twist operator is
calculated as
\begin{eqnarray}
\langle \Psi |z |\Psi \rangle
=
(-1)^{N+1}
\exp\left[
\mathrm{i}\gamma - \xi_{loc}^2/N +\mathcal{O}(1/N^2)
\right],
\label{eq: expectation value of the twist operator 2}
\end{eqnarray}
where the Berry phase (Zak's phase) $\gamma$ 
is given by
a line integral of the gauge field $A(k)$
over the 1D Brillouin zone (Bz)
\cite{Zak89, Kingsmith93, Resta94, Rem1},
\begin{eqnarray}
\mathrm{i}A_x(k)
&:=&
\langle v_-(k) |
\frac{\mathrm{d}}{\mathrm{d}k}
| v_-(k) \rangle,
\nonumber \\
\gamma 
&:=&
\mathrm{i}\int_0^{2\pi}\mathrm{i}A_x(k)\,  \mathrm{d}k\, .
\label{eq: the Berry phase}
\end{eqnarray}
For the Fermi-sea $|\Psi\rangle$ derived from 
the Hamiltonian (\ref{eq: def 2-band 1D hamiltonian in k-space}),
$\gamma$ is simply equal to half of the solid angle sustained by
a loop defined by $\boldsymbol{R}(k)$ in $\boldsymbol{R}$-space
\cite{Berry84, Ryu02}.
(Fig.\ \ref{fig: bloch_sphere})
On the other hand,
the $\mathcal{O}(1/N)$ correction to
$\ln \langle \Psi |z |\Psi \rangle$ is real
and given by the integral of the quantum metric 
$g_{xx}(k)$ over the Bz \cite{Marzari97},
\begin{eqnarray}
g_{xx}(k)
&:=&
\mathrm{Re}\,
\langle \partial_{k} v_- |
\partial_{k} v_- \rangle
-
\langle \partial_{k} v_- | v_- \rangle
\langle v_-|
\partial_{k} v_- \rangle,
\nonumber \\
\xi_{loc}^2
&:=&
\pi
\int_0^{2\pi}
g_{xx}(k)\, \mathrm{d}k\,.
\label{eq: the quantum metric}
\end{eqnarray}
The localization length $\xi_{loc}$ plays a similar role
to $\xi_{corr}$ and
is known to be related to 
most localized Wannier states in an insulating phase.
\cite{Marzari97}

\subsection{Truncated correlation matrix and its zero modes} 

We next partition the system into two 
parts, 
$A=\{x\, |\, x=1,\ldots,N_A\}$
and
$B=\{x\, |\, x=N_A+1,\ldots,N_B\}$
with $N_A+N_B=N$,
and ask, with the von Neumann entropy $S_A$,
to what extent these two subsystems are entangled.
Instead of directly tracing out the subsystem $B$
following the definition (\ref{eq: def entanglement entropy}),
we can make use of correlation matrix
$
C_{\iota\lambda}(x-y)
:=
\langle c_{(x,\iota)}^{\dag} c_{(y,\lambda)}^{\ } \rangle
$ as shown in Ref.\ \cite{Peschel02}.
From the entire correlation matrix,
we extract the submatrix
$\{C_{\iota\lambda}(x-y)\}_{x,y\in A}$ where $x$ and $y$ are restricted 
in the subsystem $A$.
The entanglement entropy is then given by
\begin{eqnarray}
S_{A}&=&
-\sum_{a}
\Big[
\zeta_{a}\ln\zeta_{a}
+
(1-\zeta_a)\ln(1-\zeta_{a})
\Big],
\label{eq: master formula for the entropy}
\end{eqnarray}
where
$\zeta_a$ are the eigen values of 
the truncated correlation matrix
$\{C_{\iota\lambda}(x-y)\}_{x,y\in A}$.

With the whole set of the eigen values 
$\{E_{\pm}(\boldsymbol{k})\}$
and eigen wavefunctions $\{v_{\pm}(\boldsymbol{k})\}$
(Eq.\ (\ref{eq : wave function for monopole}))
in hand,
the correlation matrix 
$
C_{\iota\lambda}(x-y)
=
N^{-1}
\sum_{k\in \mathrm{Bz}}
e^{-\mathrm{i}k(x-y)}
C_{\iota\lambda}(k)
$
is calculated exactly as
\begin{eqnarray}
C_{\iota\lambda}(k)
&=&
\frac{1}{2}
\big[
n^{\mu}(k)\sigma_{\mu}
\big]_{\iota\lambda},
\label{eq: corr matrix in mom space}
\end{eqnarray}
where we have introduced an ``on-shell'' four-vector 
$n^{\mu}$ by
$
n^{\mu}=(1,\boldsymbol{R}/R)
$.
It should be noted that a set of Hamiltonians can
share the same ground state wavefunction and thus
the same correlation matrix.

The basic idea we will use to discuss the entanglement
entropy is to think that the correlation matrix $C(x-y)$
defines a 1D ``Hamiltonian'' with PBC.
This ``Hamiltonian'' (let us call it 
the correlation matrix Hamiltonian or the $\mathcal{C}$-Hamiltonian
for simplicity)
has the same set of eigen wave functions as the original Hamiltonian
but all the eigen values are given by either 1 or 0.
The range of hopping elements in the generated system
is order of the inverse gap of the original Hamiltonian.
I.e., if there is a finite gap,
the $\mathcal{C}$-Hamiltonian is local (short-ranged).

Now, all we need to know
is what energy spectrum
the $\mathcal{C}$-Hamiltonian will have 
when we cut it into two parts, defined by $A$ and $B$.
This is the same question asked in Ref.\ \onlinecite{Ryu02},
in which a criterion to determine the existence of zero-energy
edge states is presented.
There are two types of eigen values in 
the energy spectrum of the truncated $\mathcal{C}$-Hamiltonian
in the thermodynamic limit $N_A \to \infty$.
Eigen values of the first type
are identical to their counterpart in the periodic (untruncated) system.
On top of it,
there appear localized boundary states whose eigen values are
located within the bulk energy gap.
Since the eigen values that belong to the bulk part
of the spectrum are either 1 or 0, they do not 
contribute to the entanglement entropy as seen from 
Eq.\ (\ref{eq: master formula for the entropy})
whereas 
the boundary modes do.

The question is then how many boundary states appear 
and with what energy when the system is truncated.
As suggested from the the KSV formula
in macroscopic polarization,
the non-vanishing Berry phase of the filled band of 
the $\mathcal{C}$-Hamiltonian implies
the existence of states localized near the boundary.
Here, note that 
the Berry phase for the generated system ($\mathcal{C}$-Hamiltonian) is identical
to that of the original system, since
the original and generated Hamiltonians
share the same set of eigen wave functions.

\subsection{Dimerized limit}
\label{subsec: limit 1}

To know the number of localized states that appear in the spectrum and
the energy eigen values thereof is, in general, a difficult task.
In this subsection,
we consider a limiting situation 
in which the localization length 
in Eq.\ (\ref{eq: the quantum metric})
is small $\xi_{corr} \ll 1$
and the band width of the energy spectrum is zero.
More precisely,
let us consider the case in which the 
correlation matrix is given by a four-vector 
$n^{\mu}$ 
(Eq.\ (\ref{eq: corr matrix in mom space}))
with
\begin{eqnarray}
&&
\boldsymbol{R}(k)=
(-\Delta \cos k,-\Delta \sin k,\xi),
\label{eq: def example 1}
\end{eqnarray}
where $\Delta,\xi \in\mathbb{R}$.
There is a family of Hamiltonians
having this correlation matrix which includes
the following ``dimerized'' Hamiltonian,
\begin{eqnarray}
\mathcal{H}
=
\sum_{x}
\left[
\sum_{\iota}
\iota \xi
c_{x\iota}^{\dag} c_{x\iota}^{\ }
+
\Delta c_{x+1,+}^{\dag} c_{x,-}^{\ }
+
\mathrm{h.c.}
\right].
\label{eq: dimerized Hamiltonian}
\end{eqnarray}
The inverse Fourier transformation of
Eq.\ (\ref{eq: def example 1})
gives
the correlation matrix in the tight-binding notation,
\begin{eqnarray}
\mathcal{C}
=
\sum_{x}
\left[
\sum_{\iota}
\frac{(R - \iota \xi)}{2R}
c_{x\iota}^{\dag} c_{x\iota}^{\ }
-
\frac{\Delta}{2R} c_{x+1,+}^{\dag} c_{x,-}^{\ }
+
\mathrm{h.c.}
\right].
\nonumber \\
\label{eq: dimerized correlation matrix}
\end{eqnarray}
This $\mathcal{C}$-Hamiltonian can be 
diagonalized for both periodic and truncated boundary conditions
by introducing the ``dimer'' operators via
$
d^{\dag}_{\pm,x+\frac{1}{2}}
=
(
c_{x,+}^{\dag}
\pm
c_{x+1,-}^{\dag}
)/\sqrt{2}.
$
(See also Appendix.)
The truncated $\mathcal{C}$-Hamiltonian has
$(N-1)$-fold degenerate eigen values
$\zeta=0,1$,
and two eigen values
$
\zeta=
(
1\pm \frac{\xi}{R}
)/2
$
that correspond to edge states.
The entanglement entropy (in the thermodynamic limit)
is then computed as
\begin{eqnarray}
\frac{1}{2}
S_{A}
&=&
-
\frac{\gamma}{2\pi}\ln \frac{\gamma}{2\pi}
-
\frac{(2\pi-\gamma)}{2\pi}\ln \frac{2\pi-\gamma}{2\pi}.
\label{eq: formula in limit 1}
\end{eqnarray}
where the Berry phase $\gamma$ for the correlation matrix 
(\ref{eq: def example 1}) is 
$
\gamma/\pi
=
1
-\xi/R.
$
In the two extreme cases,
$\xi=0$ and $\xi\to \pm \infty$,
we have
$
S_{A}(\xi=0)=2\ln 2
$
and
$
S_{A}(\xi\to \pm \infty)=0
$,
respectively.
The entanglement entropy in the present case
is a convex function with respect to $\gamma \in [0,2\pi]$
and the maximum is achieved when $\gamma=\pi$ whereas
two minima are located at $\gamma=0,2\pi$.

\subsection{Case of $\gamma=\pi$ with chiral symmetry}
\label{subsec: limit 2}

Although
the formula (\ref{eq: formula in limit 1})
clearly shows the relation between 
the Berry phase and the entanglement entropy
in a specific limit,
it is rather difficult to extend 
Eq.\ (\ref{eq: formula in limit 1}) to more
generic situations.
However
if we impose a discrete symmetry
implemented by a unitary particle-hole transformation, 
so-called chiral symmetry, 
on the $\mathcal{C}$-Hamiltonian,
it is possible to make a precise prediction for the number of 
boundary states that has an eigen value $\zeta=1/2$,
following the same line of discussions in Ref.\ \onlinecite{Ryu02}.

When the system respects the chiral symmetry,
we can find a unitary matrix that anti-commutes
with the one-particle Hamiltonian.
For this case, $\boldsymbol{n}(k)$ is restricted to lie
on a plane cutting the origin in $\boldsymbol{R}$-space, 
which in turn means that
the Berry phase for the lower band of $\mathcal{H}$
is equal to $n \pi$ ($n \in \mathbb{N}$).
(Fig.\ \ref{fig: bloch_sphere})

When $n$ is odd,
we can show that there are
at least a pair of boundary modes at $\zeta=1/2$,
one of which is localized at the left end and the other at the right.
\cite{Ryu-unpublished}
(The system with $\gamma=\pm \pi$ is, in a sense, ``dual'' to 
that with the vanishing Berry phase where there is no boundary state.
See Appendix.)
Basically, this is because, 
when $n$ is odd,
it is always possible to deform the $\mathcal{C}$-Hamiltonian
into a ``reference'' one 
without closing the bulk energy gap and without changing
the Berry phase.
The reference $\mathcal{C}$-Hamiltonian is similar to 
the dimerized example 
(\ref{eq: dimerized correlation matrix})
for which one can exactly show 
the existence of $n$ pairs of edge modes at $\zeta=1/2$.
In the course of deformation, 
the edge modes present in the reference $\mathcal{C}$-Hamiltonian
can move away from $\zeta=1/2$. However,
due to the chiral symmetry, the edge modes can escape from $\zeta=1/2$
only in a pair wise fashion, i.e., 
an edge state localized on the left/right
must always be accompanied by the one localized on the same end
and with the opposite eigen value with respect to $\zeta=1/2$.
When $n$ is odd,  
a pair of edge modes ( one for each end )
cannot have its partner and hence we are left with at least one edge
mode per boundary located 
exactly at $\zeta=1/2$.
See Ref.\ \cite{Ryu02} for more detailed discussions.

Then, the lower bound of the entanglement entropy is given by
\begin{eqnarray}
S_{A}&\ge&
-
\ln \frac{1}{2}
-
\ln \frac{1}{2}
=
2\ln 2.
\end{eqnarray}
This lower bound is equal to the entanglement entropy
contained in a dimer for each end of the original model,
which is consistent with the fact that 
the origin of the boundary states discussed above can be traced back to 
dimers in the reference Hamiltonian to which a given target
Hamiltonian is adiabatically connected.

There can be other contributions from boundary states that are not 
connected to a dimer in the above sense. 
Indeed, as we will explicitly demonstrate below,
this kind of boundary modes proliferate
as we approach a quantum critical point 
whose number grows as 
$\sim \mathcal{A}(c/6)\ln \xi_{corr}/a$,
and finally 
gives rise to the logarithmic divergences at the critical point.
\cite{Calabrese04}

Note also that our discussion here does not apply gapless systems
since 
the matrix elements of the $\mathcal{C}$-Hamiltonian are long-ranged
in this case.

\begin{figure}
\begin{center}
\includegraphics[width=4cm,clip]{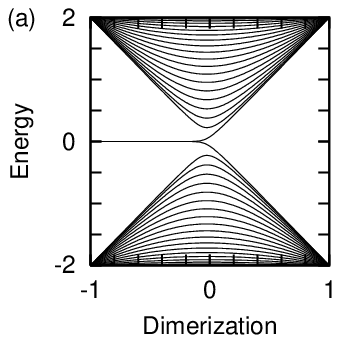}
\includegraphics[width=4cm,clip]{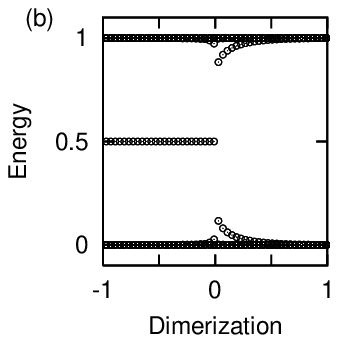}
\\
\includegraphics[width=4cm,clip]{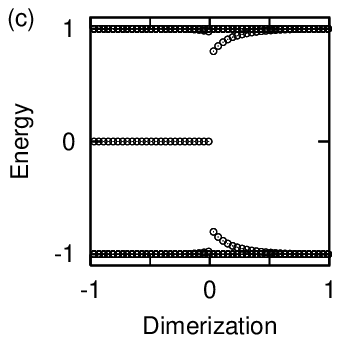}
\includegraphics[width=4.2cm,clip]{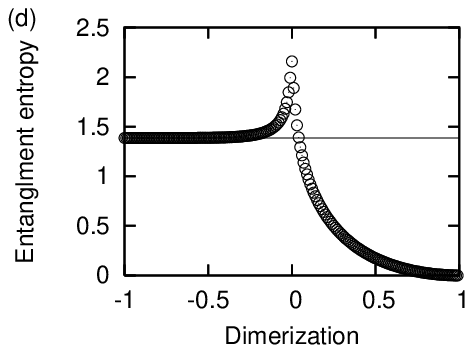}
\caption{
\label{fig: ssh edge}
The energy spectra of (a)
the Hamiltonian $\mathcal{H}$ with open ends, 
(b)the truncated correlation matrix $\mathcal{C}$,
and (c) the matrix $\mathcal{S}$
(see Sec.\ \ref{sec: connection to  a kink operator})
as a function of
the dimerization parameter $\phi\in [-1,1]$
for the SSH model.
Both energy and dimerization are measured in the unit of 
the hopping amplitude, $t$.
(d)
The entanglement entropy of the SSH model.
}
\end{center}
\end{figure}

\subsection{Example : the Su-Schrieffer-Heeger model}

As an example,
let us look at a situation
in which 
two phases with the Berry phase
$\gamma=\pi$ and 0 are connected by
a quantum phase transition point.
Physically, such an example is provided by
the Su-Schrieffer-Heeger 
(SSH) model for a chain of polyacetylene.
The 1D tight-binding Hamiltonian
for the SSH model for a chain of polyacetylene is given by
$
\mathcal{H}
=
\sum_{i=1}^{N_i}
t
\big(
-1+(-1)^i\phi_i
\big)
\big(
c_i^{\dag}c_{i+1}^{\ }
+
\mathrm{h.c.}
\big)
$
\cite{Heeger88}
where 
$\phi_{i}$ represents dimerization
at the $i$-th site, and
an alternating sign of the hopping elements
reflects dimerization between the carbon atoms
in the molecule.
Here, we treat the lattice in a classical fashion
and neglected its elastic (kinetic) energy.
Taking $\phi_i=\phi=\mathrm{const.}$, $t=1$,
and defining a spinor at $x=2i-1$ by
$
\boldsymbol{c}_{x}
=
\left(
c_i,c_{i+1}
\right)^{\mathrm{T}},
$
the Hamiltonian can be written as
($N=N_i/2$)
\begin{eqnarray}
\mathcal{H}
&=&
\sum_{x=1}^{N}
\boldsymbol{c}^{\dag}_{x}
\left(
\begin{array}{cc}
 & -(1+\phi)\\
-(1+\phi) &
\end{array}
\right)
\boldsymbol{c}^{\ }_{x}
\nonumber \\
&&
-
\boldsymbol{c}^{\dag}_{x}
\left(
\begin{array}{cc}
 & 0\\
1-\phi  &
\end{array}
\right)
\boldsymbol{c}^{\ }_{x+1}
+
\mathrm{h.c.}
\end{eqnarray}
Under the PBC, the SSH Hamiltonian can be
diagonalized as Eq.\ (\ref{eq: def 2-band 1D hamiltonian in k-space})
with
$
R_{x}(k) =-1-\phi -(1-\phi)\cos k
$,
$
R_{y}(k) =(-1+\phi)\sin k
$,
$
R_z(k)=0.
$

For $\phi \in [-1,0)$,
the Berry phase is given by $\gamma=\pi$
whereas for $\phi \in (0,1]$ $\gamma=0$.
These two phases are separated by a quantum phase
transition at $\phi=0$.
Following the discussion in Ref.\ \cite{Ryu02},
there is at least pair of boundary states
for $\phi \in [-1,0)$
when we terminated the system.
Indeed, for the numerically computed 
energy spectrum of the SSH model with open ends
(Fig.\ \ref{fig: ssh edge}-(a) ) for $\phi\in[-1,+1]$,
there is a pair of edge states in the bulk energy gap
when
$\phi \in [-1,0)$.

The entanglement entropy is
calculated by diagonalizing the $\mathcal{C}$-Hamiltonian.
The energy spectrum of the $\mathcal{C}$-Hamiltonian
with open ends is shown in Fig.\ \ref{fig: ssh edge}-(b).
Again, there is a pair of boundary states for $\phi\in[-1,0)$
and for this case, $S_A$ is bounded from below as 
$S_A \ge 2 \ln 2$. ( Fig.\ \ref{fig: ssh edge}-(d))
When we approach the transition point $\phi=0$,
some bulk eigen values turn into the boundary eigen values
and they give rise to extra contributions other than
the zero-energy boundary states.
Similar behavior of the entanglement entropy is discussed
for the quantum Ising chain in transverse magnetic field,
where the $2\ln 2$ entropy originates from 
a Schr\"odinger cat state composed of 
all spin up and down configurations.

\section{Connection to  a kink operator}
\label{sec: connection to  a kink operator}

We have seen that
bipartitioning the system  
corresponds to 
an introduction of a sharp ``boundary'' (interface).
In this section,
we will realize it by a non-local operator,
a kink operator 
\begin{eqnarray}
\eta
&:=&
\exp\left[
\mathrm{i}\sum_{x} \varphi(x) n_{x}
\right],
\quad
\eta^{\dag}=\eta^{-1},
\end{eqnarray}
where 
\begin{eqnarray}
\varphi(x)&:=&
\left\{
\begin{array}{ll}
0,   & x\in A, \\
\pi, & x \in B.
\end{array}
\right. 
\end{eqnarray}
The geometric mean of this kink operator 
is the twist operator.
\cite{Shindou05}

The kink operator attaches a phase factor $\varphi(x)$ 
for the fermion operators at site $x$,
\begin{eqnarray}
\eta^{\dag} c_{x\iota}^{\ } \eta
=
e^{+\mathrm{i}\varphi(x)} c_{x\iota},
\quad
\eta^{\dag} c_{x\iota}^{\dag} \eta
=
e^{-\mathrm{i}\varphi(x)} c_{x\iota}^{\dag}.
\label{eq: phase attachment}
\end{eqnarray}
Thus, if we introduce the reduced density operator through
\begin{eqnarray}
\tilde{\rho}_{A}
&:=&
\frac{1}{2}
\left[
\eta |\Psi\rangle\langle \Psi|\eta^{\dag}
+
|\Psi\rangle\langle \Psi|
\right],
\end{eqnarray}
the matrix elements 
$\mathrm{tr}\, \big[
c_{x,\iota}^{\dag} c_{y,\lambda}^{\ } \tilde{\rho}_A
\big]$
are vanishing whenever $x\in A$ and $y\in B$ and vice versa,
whereas they coincide with the correlation matrix
$C_{\iota\lambda}(x-y)$
when $x,y \in A$.
Unlike $\rho_A$,
the matrix elements 
$\mathrm{tr}\, \big[ c_{x,\iota}^{\dag} c_{y,\lambda}^{\ }
\tilde{\rho}_A \big]$
are non-zero even for the $B$ subsystem.
This ``padding'' does nothing however.

In the following, we will discuss
the expectation value of the kink operator 
$\langle \Psi| \eta | \Psi \rangle$ with respect to 
a given ground state wave function $|\Psi\rangle$
which is related to
the expectation value of $\tilde{\rho}_A$ as
$
\langle \Psi | \tilde{\rho}_A |\Psi \rangle 
=
\frac{1}{2}
\left[
|\langle \Psi| \eta | \Psi \rangle|^2
+1
\right]$.
As we will see the vanishing of 
$\langle \Psi| \eta | \Psi \rangle$
is closely related to a $\ln 2$ contribution
to $S_A$ discussed in the previous section.
This can be understood intuitively as follows.
Classical wave functions can be written as a product state
and are rather insensitive to the kink operator. Thus,
the ground state with the kink operator inserted
$\eta |\Psi \rangle$ has a large overlap with the original ground state
$|\Psi \rangle$.
On the other hand,
the kink operator destroys dimers
if the Berry phase of the ground state is $\pi \times (\mbox{odd integer})$.
As a consequence the overlap $\langle \Psi|\eta |\Psi \rangle$ is very
small in this quantum phase, which in turn suggests that quasi-particles
that constitute the continuum spectrum above the ground state 
can be interpreted as a kink created by $\eta$.
Thus the kink operator is capable of distinguishing the quantum phases with 
different entanglement properties.

To put the above statement in a quantum information perspective,
remember the reduced density matrix $\tilde{\rho}_A$ is in general
in a mixed state:
\begin{eqnarray}
\tilde{\rho}_A =
\sum_{n} p_{n} |\Psi_n \otimes 0\rangle
 \langle \Psi_n \otimes 0|,
\end{eqnarray}
where $|\Psi_n \rangle$ belongs to the subsystem $A$,
and $\sum_n p_n =1$.
When the wavefunction $|\Psi\rangle$ happens to be a 
completely entanglement-free, product state,
$|\Psi\rangle = |\Psi_{A}\rangle \otimes |\Psi_{B}\rangle$,
the reduced density matrix $\tilde{\rho}_A$ is in a pure state,
i.e.,
$p_{n\neq 1}=0$, $p_1=1$, $|\Psi_1 \rangle = |\Psi_A \rangle$.
On the other hand, when
$|\Psi\rangle$ is highly entangled,
taking partial trace over the $B$ subsystem generates 
many pure states $|\Psi_n\rangle$ with non-zero weight $0< p_n < 1$.
How far a given state $|\Psi_n\rangle$ from a product state can then 
be measured by taking the expectation value of the reduced density 
matrix $\tilde{\rho}_A$:
\begin{eqnarray}
\langle \Psi |
\tilde{\rho}_A  |\Psi \rangle
=
\sum_{n} p_{n} 
\langle \Psi 
|\Psi_n \otimes 0\rangle
 \langle \Psi_n  \otimes 0|\Psi \rangle.
\end{eqnarray}
Clearly, it is equal to one when 
$|\Psi_n\rangle$ is a product state whereas
it is expected to be less than one
for entangled states.

In the following subsections, we will establish that
in an insulating phase
the expectation value of the kink operator
is zero in the thermodynamic limit
when the Berry phase is 
$\pi\times (\mbox{odd integer})$
whereas it is finite otherwise.

\subsection{Expectation value of the kink operator
as a determinant}
\label{subsec: the expectation value of the kink operator as a determinant}

The computation of the expectation value of the kink operator
for a Fermi-Dirac sea
$
|\Psi\rangle
=
\prod_{k \in \mathrm{Bz}}
\alpha_{-,k}^{\dag}
|0\rangle
$ 
goes as follows.
In the momentum space, the phase attachment transformation
(\ref{eq: phase attachment})
reads
\begin{eqnarray}
\eta^{\dag}\boldsymbol{c}_{k}^{\ }\eta^{\ }
=
\sum_q f_q \boldsymbol{c}_{k-q},
\quad
\eta^{\dag}\boldsymbol{c}_{k}^{\dag}\eta^{\ }
=
\sum_q f^{*}_q \boldsymbol{c}^{\dag}_{k-q}.
\end{eqnarray}
where we introduced the Fourier components of $e^{\mathrm{i}\varphi(x)}$ by
\begin{eqnarray}
e^{\mathrm{i}\varphi(x)}
=
f(x)
=
\sum_{q\in \mathrm{Bz}} f_q e^{\mathrm{i}q x},
\end{eqnarray}
with
$q=2\pi n_q/N$ ($n_q\in \mathbb{N}$)
and
\begin{eqnarray}
f_q&=&
2
\frac{1-e^{-\mathrm{i} \pi n_q} }
{1- e^{-\mathrm{i} 2\pi n_q/N }}
\nonumber \\
&=&
\left\{
\begin{array}{ll}
\displaystyle
\frac{4}
{1- e^{-\mathrm{i} 2\pi n_q/N}}, & n_q = 1,3,\ldots, N-1,\\
\displaystyle 0, & n_q = 0,2,\ldots, N-2. \\
\end{array}
\right.
\label{eq: fourier component fq}
\end{eqnarray}
In a basis that diagonalizes the Hamiltonian,
\begin{eqnarray}
\eta^{\dag}
\boldsymbol{\alpha}_{k}^{\ }\eta^{\ }
=
\sum_{k'}
\boldsymbol{S}_{k,k'}^{\dag}
\boldsymbol{\alpha}_{k'},
\quad
\eta^{\dag}
\boldsymbol{\alpha}_{k}^{\dag}\eta^{\ }
=
\sum_{k'}
\boldsymbol{\alpha}_{k'}^{\dag}
\boldsymbol{S}_{k,k'}^{\ }
\end{eqnarray}
where a $2N\times 2N$ matrix $\boldsymbol{S}_{(k\iota)(k'\lambda)}^{\ }$
is given by
\begin{eqnarray}
\boldsymbol{S}_{(k\iota)(k'\lambda)}^{\ }
&=&
\sum_q f^{*}_q 
\left[
v^{\dag}(k-q) v(k) 
\right]_{\iota\lambda}
\delta_{k-q,k'},
\end{eqnarray}
and
$v^{\dag}(p)=
\big(
v^{\dag}_{+}(p),
v^{\dag}_{-}(p)
\big)
$.

The expectation value of the kink operator with respect to
$|\Psi\rangle$ is then represented as the determinant of
$N\times N$ matrix $\boldsymbol{S}^{\ }_{(k-)(k'-)}$,
\begin{eqnarray}
\langle \Psi|\eta|\Psi\rangle &=&
\mathrm{det}\,
\left[
\boldsymbol{S}^{\ }_{(k-)(k'-)}
\right].
\label{eq: expectation value of eta as a determinant}
\end{eqnarray}
If we define the ``hopping'' elements $t_{p,q}$
through
\begin{eqnarray}
t_{k,k-q}&:=&
\left[ v^{\dag}(k) v(k-q) \right]_{--},
\end{eqnarray}
the matrix $\boldsymbol{S}_{(k-)(k'-)}$ in 
Eq.\ (\ref{eq: expectation value of eta as a determinant})
can be represented by a tight-binding Hamiltonian as,
\begin{eqnarray}
\mathcal{S}
&= &
\sum_{k,k'}
a_{k}^{\dag}
\boldsymbol{S}_{(k-)(k'-)}
a_{k'}^{\ }
\nonumber \\
&=&
\sum_k
\sum_q
f_q t^{\ }_{k,k-q}
a_{k}^{\dag}a_{k-q}^{\ },
\end{eqnarray}
where $a_{p}^{\dag}$ ($a_{p}^{\ }$) represents
a fermionic creation (annihilation) operator defined for $p\in \mathrm{Bz}$.
This Hamiltonian can be interpreted as describing 
a quantum particle hopping on a 1D lattice.
Note that the gauge field 
$A_x(k)$ and the metric $g_{xx}(k)$
are related to 
the phase and the amplitude of the 
nearest neighbour hopping elements 
$t^{\ }_{k,k-2\pi/N}$, respectively.
The hopping matrix $t^{\ }_{k,k-q}$
is generically non-local.
Also, since the kink operator introduces
a sharp boundary in the real space,
the dual Hamiltonian is highly non-local
in $k$-space.

It is evident from 
Eq.\ (\ref{eq: expectation value of eta as a determinant})
that the vanishing of $\langle \Psi| \eta |\Psi \rangle$ 
is equivalent to existence of 
zero modes in the spectrum of the $\mathcal{S}$-Hamiltonian.
As we will see below,
the spectrum of the $\mathcal{S}$-Hamiltonian is pretty much
similar to that of the $\mathcal{C}$-Hamiltonian:
away from a critical point,
the spectrum is gapped and
all the eigen values are close to either $+1$ or $-1$, 
except a few eigen values in the gap that reflect
the Berry phase if it is non-trivial.
If the Berry phase is $\pi\times \mbox{(odd integer)}$,
there are exact zero energy eigen modes.
When we approach a critical point,
eigen values proliferate around zero energy.
Roughly speaking,
the entanglement entropy 
takes into account the distribution
of \textit{all} the eigen values of $\mathcal{S}$,
whereas the kink operator only takes into
account the products of all the eigen values.

\subsection{''Chiral symmetry'' and ``time-reversal symmetry''}
The $\mathcal{S}$-Hamiltonian has a chiral symmetry.
It directly reflects our
bipartitioning the original system
and
has nothing to do with
the chiral symmetry in the original system.
Indeed, from Eq.\ (\ref{eq: fourier component fq}),
one can see that 
$a_{k}$ with $k$ odd/even are connected to 
$a_{k'}$ with $k'$ odd/even only.
All the eigen states in $k$-space 
are connected to
their partner with the opposite energy
via 
\begin{eqnarray}
a_k 
&\to&
a'_k =
(-1)^{\mathrm{i}\pi n_k}a_k,
\quad
k=\frac{2\pi n_k}{N}.
\end{eqnarray}
which in turn means in the real space
\begin{eqnarray}
a_x &\to&
a_{x+N_A}
=
a'_x.
\end{eqnarray}

\begin{figure}
\begin{center}
\unitlength=10mm
\begin{picture}(4,4)(-2,-1)

\put(-4,1){\vector(1,0){8}}
\put(3.5,1.2){$q$}
\put(2.5,0.5){$+\pi$}
\put(-3.5,0.5){$-\pi$}
\put(0.1,0.5){$0$}
\put(0.1,2.1){$+\pi/2$}
\put(0.1,-0.4){$-\pi/2$}
\put(0,-0.8){\vector(0,1){3.5}}
\put(-1,2.5){$\mathrm{arg}\,f_q$}

\thicklines

\put(-3,1){\line(3,-1){3}}
\put(0, 0){\line(0,1){2}}
\put(0, 2){\line(3,-1){3}}

\end{picture}
\caption{
$
\mathrm{arg}\,f_q=
\mathrm{arg}\,
\left(
4
\frac{1}{1- e^{-\mathrm{i}q}}
\right)
=
-
\mathrm{arg}\,
\left(
1- e^{-\mathrm{i}q}
\right)
$
for $n_q=1,3,5,\cdots,N-1$.
\label{fig: arg f}
}
\end{center}
\end{figure}
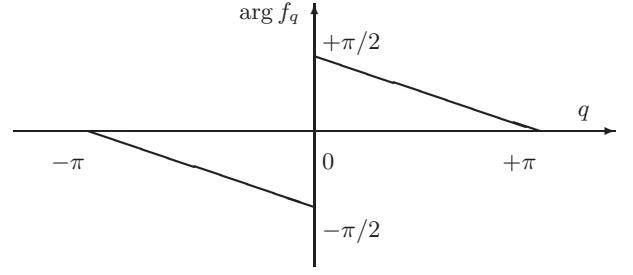

When the original system respects the
chiral symmetry (not to be confused with the chiral symmetry above),
all the single particle wave function $\psi(k)$ of $\mathcal{S}$
in $k$-space can be taken to be real
by a suitable rotation in $\boldsymbol{R}$-space.
[However when the Berry phase is $\gamma=\pi \times \mathrm{integer}$,
this comes with a price to have a Dirac string that intersect 
$\boldsymbol{R}(k)$.]
The ability of taking all ``hopping'' elements $t_{k,k'}$ 
to be real induces an additional 
``time-reversal symmetry'' to
$\mathcal{S}$-Hamiltonian;
the phase associated with $f_q$ can be removed
by a simple gauge transformation,
\begin{eqnarray}
a_k &\to & 
b_k = 
e^{+\mathrm{i}k/2-\mathrm{i}k N_A/2}a_k.
\end{eqnarray} 
[See Fig.\ \ref{fig: arg f}.]
Thus, we can take
all the matrix elements $f_q t_{k,k-q}^{\ }$
in the $\mathcal{S}$-Hamiltonian to be real.
Furthermore, when we go back to the real space, this ``time-reversal''
invariance implies a parity symmetry with respect to
an inversion center $x_0=-N_A/2+1/2$.
To see this, we first note that
all the one-particle eigen states of $\mathcal{S}$ 
can be taken real in the basis $\{b_p^{\dag},b^{\ }_p\}$;
the $\mathcal{S}$-Hamiltonian can be diagonalized as
$
\mathcal{S}
=
\sum_{n} \epsilon_{n} d^{\dag}_{n}d_{n}^{\ }
$ with
\begin{eqnarray}
b^{\ }_{p}= \sum_{n} \phi_{n}(p)d^{\ }_n,
\quad
b^{\dag}_{p}= \sum_{n} \phi_{n}(p)d_n^{\dag },
\end{eqnarray}
where $\phi_{n}(p)$ is a eigen wavefunction which is real.
Since the basis $\{a_x^{\dag},a_x^{\ }\}$  and 
$\{d_n^{\dag},d_n^{\ }\}$ are related through
\begin{eqnarray}
a_x 
&=&
\sum_n
\frac{1}{\sqrt{N}}\sum_{k} 
e^{\mathrm{i}k(x-1/2+N_A/2)}
\phi_n(k) 
d_n,
\end{eqnarray}
the real space eigen wavefunctions $\psi_n(x)$ in the basis
$\{a_x^{\dag},a_x^{\ }\}$
are given by
\begin{eqnarray}
\psi_n(x)=
\frac{1}{\sqrt{N}}\sum_{k} 
e^{\mathrm{i}k(x-1/2+N_A/2)}
\phi_n(k),
\end{eqnarray}
from which one can see $\psi_n(x)$ satisfies
\begin{eqnarray}
[\psi_n(x)]^{*}&=&
\psi_n(-x+1-N_A).
\end{eqnarray}
I.e., the wave function amplitude
is parity symmetric with respect to
$x_0 = -N_A/2+1/2$.

%
%
%
%
%
%
%

This time-reversal symmetry, which is 
plays an important role for the vanishing
of the expectation value of the kink operator.
Indeed, it is this symmetry which guarantees
existence of zero-modes of $\mathcal{S}$.

\subsection{Existence of zero-modes}

The argument that tells us
the existence of zero modes for the $\mathcal{S}$-Hamiltonian
is somewhat similar to the ``proof'' of the existence
of zero modes for the $\mathcal{C}$-Hamiltonian in that we consider a adiabatic 
change of the Hamiltonian.
The major difference comes from the fact that the chiral symmetry 
in the $\mathcal{S}$-Hamiltonian is implemented as a kind of
time-reversal symmetry as we discussed before.

We first establish that there is a pair of zero modes for $\mathcal{S}$
when we take $|\Psi\rangle$ as the ground state of 
the dimerized Hamiltonian (\ref{eq: def example 1})
with the chiral symmetry.
The hopping elements in $\mathcal{S}$ are computed
from the overlap of the Bloch wave functions as
\begin{eqnarray}
\langle v_{\pm}(p)|
v_{\pm} (q) \rangle 
\hphantom{AAAAAAAAAAAAAAAAA}
&&
\nonumber \\
=
\frac{1}{2R(R\mp R^3)}
\Big[
\Delta^2 e^{\mathrm{i}(p-q)}
+
R^2 \mp 2 R \xi+\xi^2
\Big].
&&
\end{eqnarray}
The $\mathcal{S}$-Hamiltonian is then diagonalized as
\begin{eqnarray}
\mathcal{S}=
\frac{1}{2R(R - R^3)}
\hphantom{AAAAAAAAAAAAAAAA}
&&
\nonumber \\
\times
\sum_{x}
\Big[
\Delta^2 f(x+1)
+
(R-\xi)^{2}f(x)
\Big]
a_x^{\dag}a^{\ }_x.
&&
\end{eqnarray}
We see that there are two mid-gap states with
energies
$\pm \xi/R$.
Especially when $\xi=0$,
there are a pair of zero energy states
localized at the interfaces.

We then change the Hamiltonian in a continuous fashion in such
a way that (i) it respects the chiral symmetry during the deformation,
and (ii) it does not cross the gap closing point 
(the origin of $\boldsymbol{R}$-space).
During this deformation, the Berry phase of the ground state
wavefunction is always kept to be $\pi$.
As already discussed, we can take all the Bloch wave functions
to be real and there is a ``time-reversal'' symmetry.

One can see that the zero modes never escape from $E=0$
as it constrained by the time-reversal symmetry, which
is nothing but the parity invariance with respect to $x_0=-N_A/2+1/2$.
First note that
since the $\mathcal{S}$-Hamiltonian in $k$-space is non-local,
it is short-ranged (quasi-diagonal) in the real space.
Thus, if we take the thermodynamic limit $N\to \infty$,
states that appear between the gap are spatially localized
near the interfaces located $x=1/2$ and $x= N_A+1/2$,
which separate the system into the two subsystems.

During the deformation, the two localized states,
which located at $x=1/2$ and $x=N_A+1/2$, respectively,
can in principle go away from $E=0$.
Due to the ``chiral symmetry'' of the $\mathcal{S}$-Hamiltonian, 
if one goes up from $E=0$, the other must
be go down.
However, if there is the ``time-reversal symmetry'',
each eigen state must be invariant under the space
inversion with respect to $-N_A/2+1/2$.
In order for the localized states to satisfy 
these two conditions, both of them must be located at 
$E=0$.


As an example, 
the spectrum of the $\mathcal{S}$-Hamiltonian
for the SSH model is presented in Fig.\ \ref{fig: ssh edge}-(c).
The spectrum is almost identical to 
that of the $\mathcal{C}$-Hamiltonian
and a pair of zero modes persists for the entire
quantum phase $\phi \in [-1,0)$.


\section{2D systems with the non-vanishing Chern number} 
\label{sec: 2D systems with the non-vanishing Chern number}

As far as we consider translational invariant systems,
the above 1D discussions still apply
to higher dimensions.
When a $d$-dimensional translational invariant system is bipartitioned
by a $(d-1)$-dimensional hyperplane, 
we can perform the $(d-1)$-dimensional
Fourier transformation along the interface.
The Hamiltonian is block-diagonal in terms of the wave number along
the interface $\boldsymbol{k}_{\parallel}$,
$
\mathcal{H}
=:
\sum_{\boldsymbol{k}_{\parallel}} \mathcal{H}(\boldsymbol{k}_{\parallel})$,
where $\mathcal{H}(\boldsymbol{k}_{\parallel})$ is a 1D Hamiltonian
for each $\boldsymbol{k}_{\parallel}$-subspace.
Then, the previous discussion applies to each 
$\mathcal{H}(\boldsymbol{k}_{\parallel})$.
As an example of a 2D two-band system,
let us consider 2D 
chiral $p$-wave superconductor ($p$-wave SC)
defined by
\begin{eqnarray}
\mathcal{H}
=
\sum_{\boldsymbol{r}}
\boldsymbol{c}_{\boldsymbol{r}}^{\dag}
\left(
\begin{array}{cc}
t & \Delta\\
-\Delta  & -t
\end{array}
\right)
\boldsymbol{c}_{\boldsymbol{r}+\hat{\boldsymbol{x}}}
+\mathrm{h.c.}
\qquad\qquad
&&
\nonumber \\
+
\boldsymbol{c}_{\boldsymbol{r}}^{\dag}
\left(
\begin{array}{cc}
t & \mathrm{i}\Delta\\
\mathrm{i}\Delta  & -t
\end{array}
\right)
\boldsymbol{c}_{\boldsymbol{r}+\hat{\boldsymbol{y}}}
+\mathrm{h.c.}
+
\boldsymbol{c}_{\boldsymbol{r}}^{\dag}
\left(
\begin{array}{cc}
\mu & 0\\
0  & -\mu
\end{array}
\right)
\boldsymbol{c}_{\boldsymbol{r}},
&&
\end{eqnarray}
where the integral index $\boldsymbol{r}$ runs over the 2D square
lattice,
$\hat{\boldsymbol{x}}=(1,0)$,
$\hat{\boldsymbol{y}}=(0,1)$,
and $t,\Delta,\mu \in \mathbb{R}$.
For simplicity, we set $t=\Delta=1$ in the following.
The chiral 
$p$-wave SC
has been discussed in the context of 
super conductivity in a ruthenate 
and paired states in the fractional quantum Hall effect.
\cite{Read00, volovik, Senthil98, goryo, morita, Hatsugai02}
There are four phases separated by three quantum critical points
at $\mu=0,\pm 4$, which are labeled by
the Chern number $Ch$ as
$Ch=0$ $( |\mu| > 4)$,
$Ch=-1$ $(-4 < \mu < 0)$, and
$Ch=+1$ $( 0 < \mu < +4)$.
The non-zero Chern number implies the IQHE in the spin transport.
\cite{Senthil98}

\begin{figure}
\begin{center}
\includegraphics[width=4cm,clip]{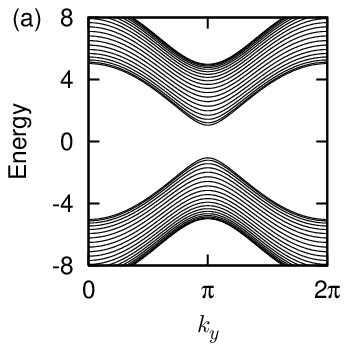}
\includegraphics[width=4cm,clip]{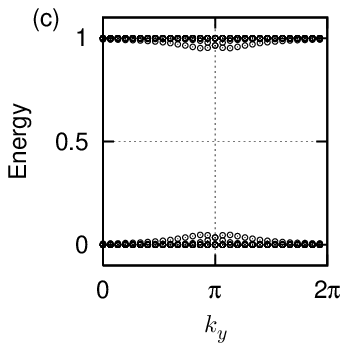}
\\
\includegraphics[width=4cm,clip]{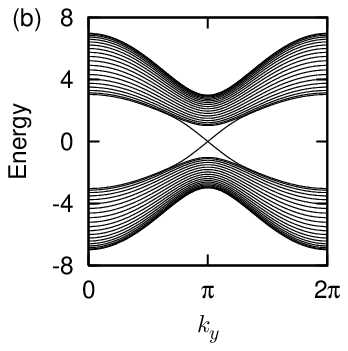}
\includegraphics[width=4cm,clip]{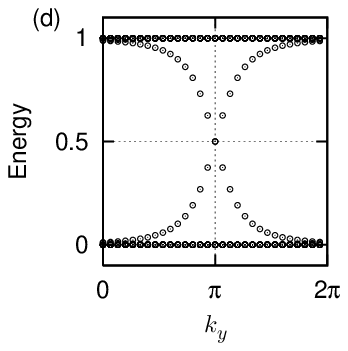}
\\
\includegraphics[width=5.5cm,clip]{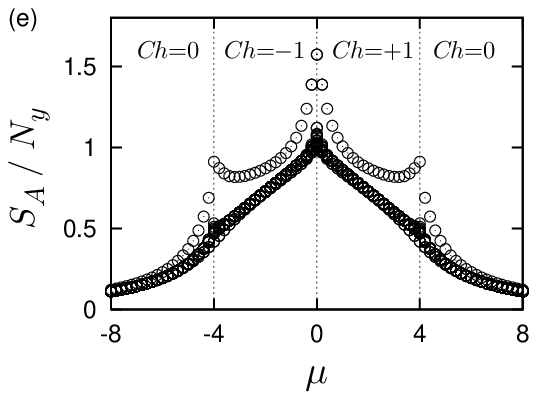}
\caption{
\label{fig: 2D p-wave edge}
The energy spectrum (measured in the unit of the hopping $t=1$)
v.s. $k_y\in [0,2\pi)$
for the 2D $p$-wave SC with boundaries.
The chemical potential is $\mu=-5$ (a) and $-3$ (b)
and $t=\Delta =1$.
The corresponding spectra of the $\mathcal{C}$-Hamiltonian
are shown in (c) ($\mu=-5$) and (d) ($\mu=-3$).
The entanglement entropy of the 2D chiral $p$-wave SC as 
a function of $\mu$ is presented in (e).n
The aspect ratio $r =N_y/N_x$ is $r =1/2, 1/3, 1/4, 1/8$ from the bottom
at $\mu=-4$.
}
\end{center}
\end{figure}

The energy spectrum of the family of Hamiltonians $\mathcal{H}(k_y)$
parametrized by the wave number in $y$-direction, $k_y$,
is given in Fig.\ \ref{fig: 2D p-wave edge}-(a),(b).
There are branches of edge states
that connects the upper and lower band
for phases with $Ch=\pm 1$.
These edge states contributes to the entanglement entropy.
The energy spectrum of the $\mathcal{C}$-Hamiltonian
with open ends is shown in Fig.\ \ref{fig: 2D p-wave edge}-(c),(d).
The corresponding entanglement entropy is also found in 
Fig.\ \ref{fig: 2D p-wave edge}-(d)
for several values of the aspect ratio $r =N_y/N_x$.
We can see that for small $r$, the entanglement entropy
shows a cusp-like behavior at quantum phase transitions
whereas for larger value of $r$,
the cusp is less eminent.

This behavior can be understood as a dimensional cross over
of the scaling behavior of the entanglement entropy 
between 1D and 2D.
For small $r$, the entropy behaves 1D-like and the
cusp is a reminiscence of the logarithmic divergent 
behavior $S_A \sim \ln N_A$ of the pure 1D case.
\cite{Holzhey94}
On the other hand, for $r$ close to unity, the entropy
exhibits a 2D behavior.
In the pure 2D limit ($r=1$),
noting that the band structure at the critical points $\mu=\pm 4$
consists of one gapless Dirac fermion,
the entanglement entropy scales as 
$S_{A} = \alpha N_y -\beta N_y/N_A$ 
where $\alpha,\beta$ is some constant.
(See Appendix \ref{app 2}.)
Notice that unlike the case of a finite Fermi surface
\cite{Wolf05,Gioev05},
$S_A/N_y$ is constant for a Dirac fermion.

An interesting and direct application of the present section
is the entanglement entropy of 
2D $d$-wave superconductors and
carbon nanotubes.
In these systems, 
different ways of bipartitioning the system 
lead to different amounts of the entanglement entropy.
\cite{Ryu02}

\section{Conclusion}
\label{sec: conclusion}

In this paper,
we have identified two types of contributions
to the entanglement entropy, i.e., 
one from the boundaries of the system created by taking
the partial trace
and the other from the bulk energy spectrum.
The contribution from the boundaries is controlled by 
the Berry phase and hence
we can make use of 
some known facts on the ``bulk-boundary correspondence''
to compute the entanglement entropy.
Especially, we have obtained the lower bound of the
entanglement entropy for 1D systems with discrete particle-hole
symmetries.
Intuitively, this means that
when the Berry phase is zero, the ground state wave function is very close to 
a simple product state, and there is not much entanglement.
Thus, ground states with non-trivial Berry phase can be
said to be more entangled in general.

Recently,
it has been revealed that 
the Berry phase manifests itself in 
the semiclassical equation of motion \cite{Sundaram99},
the density of states \cite{Xiao05}, and
the anomalous Hall effect, etc.
One can put the Berry phase correction to the entanglement
entropy in the catalog.

One of the main massages of this paper 
is the superiority of the entanglement entropy
to conventional correlation functions of local operators
to describe quantum phases.
Indeed, we clarified that the entanglement entropy
is related to non-local operators ;
the twist operator and kink operator.
The bulk contribution to the entanglement entropy is related to
the localization length (correlation length) 
which is 
the real part of the logarithm of the expectation 
value of the twist operator
and can be expressed by
the quantum metric \cite{Marzari97}.
On the other hand,
the edge contribution is tied with the imaginary part 
and to the Berry phase.
[See Eqs.\ 
(\ref{eq: expectation value of the twist operator})
to
(\ref{eq: the quantum metric}).] 
We have also made a connection between the 
entanglement entropy and the kink operator.
It is known that several phases of 1D strong correlated systems
(such as the Haldane phase)
can be described by these non-local operators.
Another connection of the entanglement entropy to a some sort of
non-local operator can be also seen in 
a recent proposal of 
a holographic derivation of the entanglement entropy.
\cite{Ryu-Takayanagi06}

Thus, the entanglement entropy can potentially be very useful 
to detect several quantum phases that need a more subtle way
of characterization than classically ordered phases.
For example, 
the entanglement entropy can be applied to
several types of spin liquid ground states
which are speculated to be described by some kind of gauge theories.
Indeed, for gapped phases of topological orders,
this direction has already been explored to some extent
\cite{Kitaev05, Levin05}.

However, in order to push this direction further,
we still need to deepen our understanding of the entanglement
entropy.
For example, extensions to multi-band systems,
especially to the case of completely degenerate bands 
might be also interesting in which we need to use
the non-Abelian Berry phase to characterize the system.
\cite{Hatsugai04}
It is also interesting to investigate if the Berry phase of quantum ground states
can be captured by
other types of entanglement measures such as
the concurrence \cite{Wootters98}.
Finally, among many other questions, we need to 
consider how we can measure the entanglement entropy
in a direct fashion. 
\cite{Klich06}



\section*{Acknowledgments}

We are grateful to 
R.\ Shindou and T.\ Takayanagi for useful discussions.
This research was supported in part
by the National Science Foundation under
Grant No.\ PHY99-07949 (SR)
and 
a Grant-in-Aid from
the ministry of Education, Culture, Sports, Science and 
Technology,
Japan (YH).

\appendix 
\section{
The dual Berry phase
}

In this appendix,
we introduce the dual Berry phase $\overline{\gamma}$ 
in the 1D two-band Hamiltonian
Eq.\ (\ref{eq: def 2-band 1D hamiltonian}).
If we impose the chiral symmetry,
in a quantum phase with $\gamma=0$ the dual Berry phase is given by
$\overline{\gamma}=-\pi$ whereas when $\gamma=-\pi$, $\overline{\gamma}=0$.
Thus a quantum phase is characterized by both $\gamma$
and $\overline{\gamma}$.


It is in spirit similar to
the dual order parameter (disorder parameter) 
in the quantum Ising spin chain.
The 1D quantum Ising model in a transverse field
has two phases : ordered and disordered phases.
It is known that the entanglement entropy is 
$S_A \ge 2\ln 2$ for the former and $S_A \ge 0$ for the latter.
On the other hand, 
they are related to each other by the Kramers-Wanier duality
and hence one may argue that they are essentially equivalent.
Why is the entanglement entropy in the ordered phase 
is larger than that in the disordered phase ?
The reason is that Kramers-Wanier duality transformation 
is a non-local transformation and does not leave original 
bipartitioning invariant.

Similarly, the duality that we will introduce momentarily 
connects two different Hamiltonians with different
Berry phase and entanglement entropy. It is possible since it 
is a transformation that changes 
the way of labeling of sites and hence
bipartitioning.

Let us first introduce dimer operators by
\begin{eqnarray}
&&
d^{\dag}_{\pm,x+\frac{1}{2}}
=
\frac{1}{\sqrt{2}}
\left(
c_{+,x}^{\dag}
\pm
c_{-,x+1}^{\dag}
\right).
\end{eqnarray}
When written in terms of the dimer operators,
the Hamiltonian 
(\ref{eq: def 2-band 1D hamiltonian})
reads
$
\mathcal{H}
=
\frac{1}{2}
\sum_{x,x'}^{\mbox{\begin{tiny}PBC\end{tiny}}}
\boldsymbol{d}^{\dag}_{x+\frac{1}{2}}
\overline{H}_{x-x'}
\boldsymbol{d}^{\ }_{x'+\frac{1}{2}}
$,
where the new hopping matrix elements 
$\overline{H}_{x-x'}$
are some function of the original ones
$H_{x-x'}$.
For simplicity,we focus on the case of particle-hole symmetric 
($t_{+}=-t_{-}$)
and translational invariant
systems.
In the momentum space, the Hamiltonian is given by
$
\mathcal{H}
=
\sum_{k \in \mathrm{Bz}}
\boldsymbol{d}_{k}^{\dag}
\overline{\boldsymbol{R}}(k)\cdot \boldsymbol{\sigma}
\boldsymbol{d}_{k}^{\ }
$
with a 3D vector $\overline{\boldsymbol{R}}(k)$ given by
\begin{eqnarray}
\overline{R}_x(k)&=&R_z(k),
\nonumber \\
\overline{R}_y(k)&=&
\sin(k) R_x(k)
+
\cos(k) R_y(k),
\nonumber \\
\overline{R}_z(k)
&=&
\cos(k) R_x(k)
-
\sin(k) R_y(k).
\end{eqnarray}
We define the dual Berry phase $\overline{\gamma}$
as the Berry phase for the dual 3D vector
$\overline{\boldsymbol{R}}(k)$,
\begin{eqnarray}
\overline{\gamma}:=
\int_0^{2\pi}\! \mathrm{d}k
\langle \overline{v}(k) |\frac{\mathrm{d}}{\mathrm{d}k}
| \overline{v}(k) \rangle
=
\int_0^{2\pi}
\mathrm{d}k\!
\frac{\overline{X}\dot{\overline{Y}}-\overline{Y}\dot{\overline{X}} }
{2 \overline{R} (\overline{R}- \overline{Z})}
.
\end{eqnarray}
where $\dot{X}=\mathrm{d}R_x(k)/\mathrm{d}k $, etc.
Rotating $\overline{\boldsymbol{R}}$ 
around $\overline{R}_y$-axis as 
$(\overline{R}_x,\overline{R}_y,\overline{R}_z) \to 
 (\overline{R}_z,\overline{R}_y,-\overline{R}_x) $,
and
noting 
$
\overline{Y}\, \dot{\overline{X}}-\overline{X}\, \dot{\overline{Y}}
=
X\dot{Y}
-
Y\dot{X}
+
X^2+Y^2
$,
the dual Berry phase is thus given by
\begin{eqnarray}
\overline{\gamma}
&=&
-\gamma
-\pi
-
\int_0^{2\pi}
\mathrm{d}k
\frac{Z}{2 R}
\end{eqnarray}
Especially if we impose the chiral symmetry,
$\boldsymbol{R}$ is restricted to $XY$-plane and thus
\begin{eqnarray}
\overline{\gamma}
&=&
-\gamma
-\pi.
\end{eqnarray}

\section{entanglement entropy
for a Dirac fermion in 2D}
\label{app 2}

In this Appendix, we estimate how much entanglement entropy
is carried by a gapless Dirac fermion.
It is an interesting question since
a Dirac fermion is just in-between of a fully gapped system
and a system with a finite Fermi surface;
for the former
the entanglement entropy satisfies the area law,
whereas for the latter
there is a log-correction to the area law.
\cite{Wolf05,Gioev05}

Unlike the case of a finite Fermi surface,
the entanglement entropy divided by $N_y$, $S_A/N_y$ is constant
for a system with Dirac fermions as can be seen as follows.
The energy spectrum close to a gap closing point in the Bz,
$\boldsymbol{k}^{(0)}$,
is linear and so is the mass gap as a function of $k_y$,
$m(k_y) \propto k_y-k^{(0)}_{y}$.
The known result for a massive 1D system tells us each $k_y$ contributes
to the entanglement entropy by $\sim \ln {m(k_y)}^{-1}$.
If the length $N_A$ of the subsystem $A$ in $x$-direction is finite,
we expect the contributions from those $k_y$ with $ m(k_y)^{-1} \ge N_A$
are given by $\sim \ln N_A$, instead.
Then, the entanglement entropy can be evaluated by summing over the
entanglement entropy for each 1D system with a fixed $k_y$,
\begin{eqnarray}
\frac{S_A }{2}
=
\sum_{k_y >0 }^{ m(k_y)^{-1} \le N_A}
\frac{2}{6}\ln m(k_y)^{-1}
+
\sum_{k_y >  0}^{ m(k_y)^{-1}\ge N_A}
\frac{1}{3}\ln N_A.
\nonumber \\
\end{eqnarray}
Converting the summation to the integral,
we see that the entanglement entropy behaves as
$S_{A} = \alpha N_y -\beta N_y/N_A$ for a single Dirac fermion
where $\alpha,\beta$ is some constant. Hence $S_A/N_y$ is finite.
This crude approximation is actually overestimating the entropy,
 but it is enough to derive essential features of the entropy.
For more detailed analysis using the entropic $c$-function, 
see Ref.\ \cite{Casini05}.



\begin{thebibliography}{99}
\bibitem{footnote1}
The base of logarithm can be taken either $e$ or $2$,
depending on literatures.
\bibitem{Rem1}
The expectation value of the twist operator
(Eq.\ (\ref{eq: expectation value of the twist operator 2}))
can be obtained in a similar fashion to that of 
the kink operator, see 
Sec.\ \ref{subsec: the expectation value of the kink operator as a determinant}.
\bibitem{Ryu-unpublished}
The similar criterion can be developed
for systems with Kramers degeneracy.
There, the time-reversal symmetry can be
in addition to the chiral symmetry.
\cite{Ryu02}
\end{thebibliography}
\end{document}